\documentclass
[showpacs,amsfonts,amssymb,oneside,notitlepage,10pt,a4paper,twocolumn,preprint,balancelastpage,pra]{revtex4}%
\usepackage{amsfonts}
\usepackage{amsmath}
\usepackage{amssymb}
\usepackage{graphicx}
\usepackage[nonscript]{accents}%
\providecommand{\U}[1]{\protect\rule{.1in}{.1in}}

\begin{document}
\title{Noncritical quadrature squeezing \\ through spontaneous polarization symmetry breaking}
\author{Ferran V. Garcia--Ferrer, Carlos Navarrete--Benlloch, Germ\'{a}n J. de Valc\'{a}rcel, and Eugenio Rold\'{a}n$^*$}
\affiliation{\small{Departament d'\`{O}ptica, Universitat de Val\`{e}ncia, Burjassot, Spain.}
\\
$^*$\small{eugenio.roldan@uv.es}}

\begin{abstract}
We discuss the possibility of generating noncritical
quadrature squeezing by spontaneous polarization symmetry breaking. We
consider first type--II frequency--degenerate optical parametric oscillators,
but discard them for a number of reasons. Then we propose a four--wave mixing
cavity in which the polarization of the output mode is always linear but has
an arbitrary orientation. We show that in such a cavity complete noise
suppression in a quadrature of the output field occurs, irrespective of the
parameter values.
\end{abstract}


\maketitle

\noindent Spontaneous rotational symmetry breaking has recently been shown theoretically
to be a resource for noncritical quadrature squeezing both in type-I
frequency--degenerate optical parametric oscillators (DOPOs) \cite{PRL08,PRA10}
and in degenerate four wave mixing $\chi^{\left(  3\right)  }$ cavities
\cite{IEEE09}, as it was previously shown for spontaneous translational
symmetry breaking in wide aperture type--I DOPOs \cite{EPL06,PRA07}. The
underlying idea is that a spontaneous symmetry breaking occurring within a
nonlinear cavity entails the existence of a canonical pair of observables: One
of them is solely driven by quantum fluctuations (such as the orientation of a
Hermite--Gauss signal mode in \cite{PRL08,IEEE09,PRA10}, or the location in
the transverse plane of a localized structure in \cite{EPL06, PRA07}) while
its canonical pair is maximally damped and, consequently, maximally
insensitive to quantum fluctuations (angular and linear momenta, respectively,
in the referred cases). We have predicted in the above cases that by this
means perfect quadrature squeezing appears in the maximally damped mode
independently of the parameter setting, hence the name \textit{noncritical}
quadrature squeezing. In the present letter we discuss on the possibility of
achieving the same result through a spontaneous polarization symmetry breaking (SPSB).

In order to achieve noncritical quadrature squeezing through SPSB the
nonlinear cavity must verify some requisites. It must posses a \textit{free
polarization parameter} (FPP), that is, a parameter of the signal's field
polarization \cite{NotaPP} not fixed by the system's dynamical equations.
Additionally, it must allow the existence of a nonzero mean field solution for the signal
field; in this way, once the threshold for the generation of the signal mode
is crossed, the random occurrence of a particular FPP will automatically
produce a SPSB. From a mathematical point of view, a system like this has a
Goldstone mode (a mode with null eigenvalue in the stability matrix
irrespective of the system parameters) reflecting the equal likeliness of any
value of the FPP, and a maximally damped mode which can be proved to be the
canonical pair of the FPP and coincides with a quadrature of the mode with
orthogonal polarization with respect to the generated one. Hence, quantum
fluctuations can freely act onto the FPP making it completely undetermined,
allowing then for the complete noise reduction of its canonical pair via the
Heisenberg uncertainty principle, that is, perfect squeezing of the
aforementioned quadrature. The above requisites can be met, at least in
principle, in both $\chi^{\left(  2\right)  }$ and $\chi^{\left(  3\right)  }$
nonlinear cavities.

Let us first consider a $\chi^{\left(  2\right)  }$ cavity. Up--conversion
processes (such as second harmonic generation) cannot help for our purposes as
the generated field appears linearly in the Hamiltonian and all its properties
are fixed by the subharmonic modes. With respect to down--conversion
processes, the only possibility is a type--II optical parametric oscillator
(OPO). In its usual configuration, type-II parametric down conversion takes
place in two modes having orthogonal linear polarizations and an
\textit{undefined relative phase} \cite{Boyd}. If in addition the
down-converted fields have the same frequency, the process just described is
equivalent to the spontaneous generation of a field with an elliptical
polarization along the $\pm45%
\operatorname{{{}^\circ}}%
$ axes having an arbitrary eccentricity and direction of rotation (see below).
Then a type--II frequency--degenerate OPO is a candidate for SPSB. In the interaction picture, the
Hamiltonian of such a system can be written as%
\begin{equation}
\hat{H}_{\mathrm{OPO}}=i\hbar\left(  \mathcal{E}_{\mathrm{p}}\hat{b}^{\dagger
}+\chi\hat{b}\hat{a}_{x}^{\dagger}\hat{a}_{y}^{\dagger}\right)  +\mathrm{H.c.}%
,\label{H0}%
\end{equation}
where $\mathcal{E}_{\mathrm{p}}$ is the pumping field amplitude, $\hat
{b}^{\dagger}$, $\hat{a}_{x}^{\dagger}$, and $\hat{a}_{y}^{\dagger}$ are the
creation operators for the pump mode and the $\mathbf{e}_{x}$ and
$\mathbf{e}_{y}$ polarized signal modes, respectively, and $\chi$ is the
nonlinear coupling constant. This Hamiltonian has the symmetry $\left(
\hat{a}_{x},\hat{a}_{y}\right)  \rightarrow\left(  \hat{a}_{x}e^{i\theta}%
,\hat{a}_{y}e^{-i\theta}\right)  $ that leaves undefined the phase difference
between the signal modes $2\theta$. This is the FPP of the system, which as
explained below, is directly related to the eccentricity of the signal field's
polarization ellipse. Associated to this symmetry there is a constant of
motion, namely the photon number difference between the signal modes $\hat
{a}_{x}^{\dagger}\hat{a}_{x}-\hat{a}_{y}^{\dagger}\hat{a}_{y}$. This ensures
that signal photons are created in pairs, and hence the two polarization modes
$\mathbf{e}_{x}$ and $\mathbf{e}_{y}$ will have exactly the same properties:
they are \textit{twin beams }whose intensity difference is potentially
perfectly squeezed \cite{Reynaud87,Heidmann87}.

Hamiltonian (\ref{H0}) is isomorphic to that of \cite{PRL08,PRA10}. There we
analyzed squeezing generation through spontaneous rotational symmetry breaking
in a type--I DOPO, in which the two signal modes had opposite orbital angular
momenta (two-transverse-mode DOPO). Hence all the results found in that system
apply for the current case. In particular, we demonstrated that: (i) for
$\mathcal{E}_{\mathrm{p}}>\gamma_{\mathrm{p}}\gamma_{\mathrm{s}}/\chi$
($\gamma_{\mathrm{p}/\mathrm{s}}$ is the cavity damping rate at the pump/signal frequency) a
steady, nonzero mean field appears at the signal frequency in the mode
$\mathbf{e}_{\mathrm{B}}=\left(  \mathbf{e}_{x}e^{-i\theta}+\mathbf{e}%
_{y}e^{i\theta}\right)  /\sqrt{2}$, an elliptically polarized mode as
explained above whose eccentricity depends on $\theta$ \cite{Guenther}%
\ (\textit{bright mode} in the following, as it is macroscopically occupied);
(ii) starting from a value dictated by the initial random fluctuations,
quantum noise makes $\theta$ diffuse; and (iii) the phase quadrature of the
mode $\mathbf{e}_{\mathrm{D}}=i\left(  \mathbf{e}_{x}e^{-i\theta}%
-\mathbf{e}_{y}e^{i\theta}\right)  /\sqrt{2}$ is perfectly squeezed (we shall
call this the \textit{dark mode} in the following, as its polarization is
orthogonal to the bright mode, and hence it is empty at the classical level).

However type--II OPOs degenerated in frequency and polarization invariant do
not seem to exist. Normally the signal modes have different frequencies
(that difference, however, can be as small as $150$\textrm{kHz} \cite{Feng03}), and the only way by which the amplification
can be made frequency degenerate is, as far as we know, by breaking the
polarization symmetry \cite{Longchambon04,Laurat05}: A birefringent plate is
introduced within the cavity, which couples the two orthogonally polarized
signal modes, thus forcing their frequency degeneracy but fixing the phase
difference between them and thus breaking the system's polarization symmetry.
Hence, given the difficulties of having frequency--degenerate type--II OPOs, we
pass to consider an alternative.

We propose a $\chi^{\left(  3\right)  }$ cavity in which SPSB can squeeze the dark
output mode. Consider an isotropic $\chi^{\left(  3\right)  }$ medium placed
inside a polarization isotropic cavity which is pumped by two copropagating
orthogonally polarized modes of frequencies $\omega_{1}$ and $\omega_{2}$ such
that $\omega_{\mathrm{s}}=\left(  \omega_{1}+\omega_{2}\right)  /2$ be close
to a cavity mode resonance $\omega_{\mathrm{c}}$. Unlike in OPOs, operation in
frequency degeneracy has been experimentally proved in this kind of systems
\cite{Vallet90}, i.e., the pump modes can be mixed to generate a signal field
with frequency $\omega_{\mathrm{s}}$ via four wave mixing. For the sake of
simplicity, we shall treat the pumping modes as classical fields and will
further ignore their depletion in interacting with the intracavity signal modes.

We write the total field at the cavity waist plane (where the $\chi^{\left(
3\right)  }$ medium is placed) as $\mathbf{\hat{E}}\left(  \mathbf{r}%
,t\right)  =\mathbf{E}_{\mathrm{p}}\left(  \mathbf{r},t\right)  +\mathbf{\hat
{E}}_{\mathrm{s}}\left(  \mathbf{r},t\right)  $,
\begin{subequations}
\label{quantum-field}%
\begin{align}
\mathbf{E}_{\mathrm{p}} &  =i\mathcal{F}G\left(  \mathbf{r}\right)
\sum_{j=1,2}\left(  \mathbf{e}_{x}\alpha_{jx}\mathbf{+e}_{y}\alpha
_{jy}\right)  e^{-i\omega_{j}t}+\mathrm{c.c.},\\
\mathbf{\hat{E}}_{\mathrm{s}} &  =i\mathcal{F}G\left(  \mathbf{r}\right)
\left[  \mathbf{e}_{x}\hat{a}_{x}\left(  t\right)  +\mathbf{e}_{y}\hat{a}%
_{y}\left(  t\right)  \right]  e^{-i\omega_{\mathrm{s}}t}+\mathrm{H.c.},
\end{align}
being $\mathcal{F}^{2}=\hbar\omega_{\mathrm{s}}/\left(  2\varepsilon
_{0}nL\right)  $ ($n$ is the refractive index and $L$ the effective cavity
length), and $G\left(  \mathbf{r}\right)  =\sqrt{2/\pi}w^{-1}\exp\left(
-r^{2}/w^{2}\right)  $ the cavity TEM$_{00}$ mode with radius $w$ (which for
simplicity we assume to be the same for pump and signal modes). Using the
properties of the nonlinear susceptibility of isotropic media and ignoring its
dispersion in the working frequency range \cite{NotaSus}, the interaction
picture Hamiltonian describing our $\chi^{\left(  3\right)  }$ cavity reads
$\hat{H}=\hat{H}_{0}+\hat{H}_{\mathrm{int}}$, where
\end{subequations}
\begin{subequations}
\begin{align}
\hat{H}_{0} &  =\hbar\delta\left(  \hat{a}_{x}^{\dagger}\hat{a}_{x}+\hat
{a}_{y}^{\dagger}\hat{a}_{y}\right)  ,\\
\hat{H}_{\mathrm{int}} &  =\frac{3}{4}\hbar g\left(  \hat{H}_{\mathrm{spm}%
}+\hat{H}_{\mathrm{cpm}}+\hat{H}_{\mathrm{fwm}}\right)  ,
\end{align}
with $\delta=\omega_{\mathrm{c}}-\omega_{\mathrm{s}}$ and $g=-8\varepsilon
_{0}l\chi_{xxxx}\mathcal{F}^{4}/\left(  \hbar\pi w^{2}\right)  $.
We assumed that the $\chi^{\left(  3\right)  }$ medium length $l$ is smaller
than the cavity Rayleigh length. In $\hat{H}_{\mathrm{int}}\ $the terms
$\hat{H}_{\mathrm{spm}}$, $\hat{H}_{\mathrm{cpm}}$, and $\hat{H}%
_{\mathrm{fwm}}$ describe self--phase modulation, cross--phase modulation, and
four--wave mixing processes, respectively; they explicitly read%
\end{subequations}
\begin{subequations}
\begin{align}
\hat{H}_{\mathrm{spm}} &  =\hat{a}_{x}^{\dagger2}\hat{a}_{x}^{2}+\hat{a}%
_{y}^{\dagger2}\hat{a}_{y}^{2},\\
\hat{H}_{\mathrm{cpm}} &  =\sum_{j=1,2}4\left(  \left\vert \alpha
_{jx}\right\vert ^{2}+\mathcal{A}\left\vert \alpha_{jy}\right\vert
^{2}\right)  \hat{a}_{x}^{\dagger}\hat{a}_{x}\nonumber\\
&  \sum_{j=1,2}4\left(  \left\vert \alpha_{jy}\right\vert ^{2}+\mathcal{A}%
\left\vert \alpha_{jx}\right\vert ^{2}\right)  \hat{a}_{y}^{\dagger}\hat
{a}_{y}\nonumber\\
&  +4\mathcal{A}\hat{a}_{x}^{\dagger}\hat{a}_{x}\hat{a}_{y}^{\dagger}\hat
{a}_{y},\\
\hat{H}_{\mathrm{fwm}} &  =\mathcal{B}\hat{a}_{x}^{\dagger2}\hat{a}_{y}%
^{2}+2\left(  \alpha_{1x}\alpha_{2x}+\mathcal{B}\alpha_{1y}\alpha_{2y}\right)
\hat{a}_{x}^{\dagger2}\nonumber\\
&  +2\left(  \alpha_{1y}\alpha_{2y}+\mathcal{B}\alpha_{1x}\alpha_{2x}\right)
\hat{a}_{y}^{\dagger2}\nonumber\\
&  +\sum_{j=1,2}4\left(  \mathcal{B}\alpha_{jx}^{\ast}\alpha_{jy}%
+\mathcal{A}\alpha_{jx}\alpha_{jy}^{\ast}\right)  \hat{a}_{x}^{\dagger}\hat
{a}_{y}\nonumber\\
&  +4\mathcal{A}\left(  \alpha_{1x}\alpha_{2y}+\alpha_{1y}\alpha_{2x}\right)
\hat{a}_{x}^{\dagger}\hat{a}_{y}^{\dagger}\mathrm{+H.c.},
\end{align}
with $\mathcal{A}=\chi_{xxyy}/\chi_{xxxx}$ and $\mathcal{B}=\chi_{xyyx}%
/\chi_{xxxx}$ that verify $2\mathcal{A}+\mathcal{B}=1$ \cite{Boyd}.

In order to preserve polarization invariance, the classical pumping fields
must necessarily have orthogonal circular polarizations as any other
polarization would privilege particular spatial directions in the transverse
plane. Consequently we take $\alpha_{1x}=\alpha_{2x}=\rho/\sqrt{2}$ and
$\alpha_{1y}=-\alpha_{2y}=i\rho/\sqrt{2}$. Then we rewrite the Hamiltonian in
the basis of circularly polarized states $\hat{a}_{\pm}=\left(  \hat{a}_{x}\mp
i\hat{a}_{y}\right)  /\sqrt{2}$ and get%
\end{subequations}
\begin{subequations}
\label{Hcirc}%
\begin{align}
\hat{H}_{0} &  =\hbar\delta\left(  \hat{a}_{+}^{\dagger}\hat{a}_{+}+\hat
{a}_{-}^{\dagger}\hat{a}_{-}\right)  ,\\
\hat{H}_{\mathrm{spm}} &  =\left(  1-\mathcal{B}\right)  \left(  \hat{a}%
_{+}^{\dagger2}\hat{a}_{+}^{2}+\hat{a}_{-}^{\dagger2}\hat{a}_{-}^{2}\right)
,\\
\hat{H}_{\mathrm{cpm}} &  =2\left(  1+\mathcal{B}\right)  \hat{a}_{+}%
^{\dagger}\hat{a}_{+}\hat{a}_{-}^{\dagger}\hat{a}_{-}\nonumber\\
&  +2\rho^{2}\left(  3-\mathcal{B}\right)  \left(  \hat{a}_{+}^{\dagger}%
\hat{a}_{+}+\hat{a}_{-}^{\dagger}\hat{a}_{-}\right)  ,\\
\hat{H}_{\mathrm{fwm}} &  =2\rho^{2}\left(  1+\mathcal{B}\right)  \left(
\hat{a}_{+}\hat{a}_{-}+\hat{a}_{+}^{\dagger}\hat{a}_{-}^{\dagger}\right)  .
\end{align}
\end{subequations}

Just as $\hat{H}_{\mathrm{OPO}}$,\ this Hamiltonian has the symmetry $\left(
\hat{a}_{+},\hat{a}_{-}\right)  \rightarrow\left(  \hat{a}_{+}e^{i\theta}%
,\hat{a}_{-}e^{-i\theta}\right)  $ with $\hat{a}_{+}^{\dagger}\hat{a}_{+}%
-\hat{a}_{-}^{\dagger}\hat{a}_{-}$ as the associated constant of motion.
Hence, whenever a nonzero mean field solution appears for the signal modes,
bright emission will take place in the mode $\mathbf{e}_{\mathrm{B}}=\left(
\mathbf{e}_{+}e^{-i\theta}+\mathbf{e}_{-}e^{i\theta}\right)  /\sqrt
{2}=\mathbf{e}_{\theta}$, i.e., the mean field will have a linear polarization
along the arbitrary $\theta$ axis \cite{Guenther} ($\mathbf{e}_{\pm}$ are the right and
left circularly polarized modes), thus breaking the polarization symmetry of
the system. Then, quantum fluctuations should induce a diffusion process in
the FPP $\theta$, allowing then for perfect noncritical squeezing in a
quadrature of the dark mode $\mathbf{e}_{\mathrm{D}}=-i\left(  \mathbf{e}%
_{+}e^{-i\theta}-\mathbf{e}_{-}e^{i\theta}\right)  /\sqrt{2}=\mathbf{e}%
_{\theta+\pi/2}$, which is crossed polarized with respect to the bright one.
Note that these linearly polarized modes are mapped onto the elliptically
polarized bright and dark modes of the type--II frequency--degenerate OPO by a quarter-wave plate \cite{Guenther}.

Fortunately, it will not be necessary to prove the above conclusions
explicitly as by taking $\mathcal{A}=\mathcal{B}=1/3$ (which applies when the
Kleinmann symmetry is verified like in nonresonant electronic response
\cite{Boyd}) Hamiltonian (\ref{Hcirc}) becomes isomorphic to that of
\cite{IEEE09}, where we already proved that for $\delta>\sqrt{3}%
\gamma_{\mathrm{s}}$ and for pump intensities $\rho^{2}$ inside the region
defined by the curves $\rho^{2}=\gamma_{\mathrm{s}}/2g$ and $\rho^{2}=\left(
2\delta+\sqrt{\delta^{2}-3\gamma_{\mathrm{s}}^{2}}\right)  /6g$, a steady,
nonzero mean field solution for the signal modes appears through a subcritical
pitchfork bifurcation. The perfect and noncritical squeezing of a quadrature
of the dark mode was also proved within the linear approximation for quantum
fluctuations. Except for the quantitative details, these results hold for
$\mathcal{A}\neq\mathcal{B}$.

Some comments are in order here. First, one may think that the non--depletion
of the pumping modes could be a too drastic approximation. But the effect of
pump depletion could consist only in introducing new bifurcations affecting
the stability of the steady state signal field. This would reduce the domain
of existence of the signal field cw state, but would not affect its
fluctuations properties where stable. Second, in \cite{PRA10} we demonstrated
for the aforementioned two-transverse-mode DOPO both the random rotation of
the bright mode and the perfect noncritical squeezing of the dark mode by
direct numerical integration of the system's quantum dynamic equations, hence
showing that these properties hold beyond the linear approximation. Finally,
in \cite{PRL08} we proved that small deviations from the perfect rotational
symmetry don't imply a large degradation of the dark mode's quadrature
squeezing, while it can fix the bright and dark mode's orientation, what seems
quite advantageous from the experimental point of view \cite{NotaParis}. These
conclusions will doubtless apply to the $\chi^{\left(  3\right)  }$ cavity we
have presented here.

Finally, note that although one of the Stokes parameters is free from quantum
fluctuations in the light exiting the cavity (the twin beams' intensity
difference), this is not enough to claim for polarization squeezing as further
conditions must be satisfied \cite{Korolkova02,Ferran07}.

In conclusion, we have theoretically demonstrated that type--II frequency--degenerate OPOs and appropriate $\chi^{\left(  3\right)  }$ cavities are
suitable for the generation of noncritical and perfect quadrature squeezing.
We have commented the problems that actual type--II frequency--degenerate OPOs
may have (either they are not exactly frequency degenerated or have a cavity
that is not polarization symmetric). We believe that the $\chi^{\left(
3\right)  }$ cavity we are proposing as an alternative can be built within the
experimental state of the art and would not present the same problems as the
$\chi^{\left(  2\right)  }$ cavity, hence being a good candidate for observing SPSB.

This work has been supported by the Spanish Ministerio de Ciencia e
Innovaci\'{o}n and the European Union FEDER through Project
FIS2008-06024-C03-01. C N-B is a grant holder of the FPU programme of the
Ministerio de Educaci\'{o}n y Ciencia.


\begin{thebibliography}{99}
\bibitem {NotaPP}The FPP might be, e.g., the eccentricity or the orientation
of the polarization ellipse. We remind that continuous variations of these
parameters correspond to rotations on the Poincar\'{e} sphere defined by the
Stokes parameters \cite{Guenther}.
\bibitem {NotaSus}The explicit form of the $\chi^{\left(  3\right)  }$ nonlinear
susceptibility tensor taking into account these considerations is $\chi
_{ijkl}\left(  \omega_{a}+\omega_{b}-\omega_{c};\omega_{a},\omega_{b}%
,-\omega_{c}\right)  =\chi_{xxyy}\left(  \delta_{ij}\delta_{kl}+\delta
_{ik}\delta_{jl}\right)  +\chi_{xyyx}\delta_{il}\delta_{jk}$ irrespective of
the frequencies involved in the particular process \cite{Boyd}.
\bibitem {NotaParis}In fact, in the type--II OPO experiment commented above
\cite{Laurat05} 4.5 \textrm{dB} of noise reduction was found in the dark
mode even though the polarization symmetry was broken as explained in the text.
\end{thebibliography}
\end{document}